\documentclass[reprint,prx,aps,longbibliography,footinbib,floatfix,nobalancelastpage,superscriptaddress]{revtex4-2}
\usepackage{amsmath}
\usepackage{multirow}
\usepackage{amsfonts}
\usepackage{amssymb}
\usepackage{mathtools}
\usepackage{graphicx}
\usepackage{subfigure}
\usepackage[dvipsnames]{xcolor}
\usepackage[colorlinks=True,citecolor=gray,linkcolor=myRed,urlcolor=gray]{hyperref}
\usepackage{hypcap}
\usepackage{yfonts}
\usepackage{bm}
\usepackage{soul}
\usepackage{ulem} 
\usepackage{dsfont}
\usepackage{braket}
\newcommand{\brakett}[2]{\ensuremath{\left\langle {#1} | {#2} \right\rangle}}

\newcommand\eq[1]{\begin{align}#1\end{align}}

\usepackage{color,xcolor, soul}

    \definecolor{darkblue}{rgb}{0,0,.65}
    \definecolor{darkgreen}{rgb}{0.3,0.9,0.3}
    \definecolor{darkorange}{rgb}{0.85,0.65,0.3}
    \definecolor{cyan1}{rgb}{0.0, 0.6, 0.6}

\definecolor{myBlue}{RGB}{31,119,180}
\definecolor{myOrange}{RGB}{255,127,14}
\definecolor{myGreen}{RGB}{44,160,44}
\definecolor{myRed}{RGB}{214,39,40}
\definecolor{myPurple}{RGB}{148,103,189}

\makeatletter
\def\p@figure{\color{myBlue}}
\def\p@equation{\color{myRed}}
\makeatother

\begin{document}

\title{Robust Correlation-Induced Localization Under Time-Reversal Symmetry Breaking}

\author{Bikram Pain}
\email{bikram.pain@icts.res.in}
\affiliation{International Centre for Theoretical Sciences, Tata Institute of Fundamental Research, Bengaluru 560089, India}
\affiliation{Nordita, Stockholm University and KTH Royal Institute of Technology, Hannes Alfv\'ens v\"ag 12, SE-106 91 Stockholm, Sweden}

\author{Sthitadhi Roy}
\email{sthitadhi.roy@icts.res.in}
\affiliation{International Centre for Theoretical Sciences, Tata Institute of Fundamental Research, Bengaluru 560089, India}
\author{Jens H. Bardarson}
\affiliation{Department of Physics, KTH Royal Institute of Technology, 106 91 Stockholm, Sweden}
\author{Ivan M. Khaymovich}
\email{ivan.khaymovich@gmail.com}
\affiliation{Nordita, Stockholm University and KTH Royal Institute of Technology, Hannes Alfv\'ens v\"ag 12, SE-106 91 Stockholm, Sweden}

\begin{abstract}
We study Anderson localization in a one-dimensional disordered system with long-range correlated hopping decaying as $1/r^{a}$ with complex hopping amplitudes that break time-reversal symmetry in a tunable fashion by varying their argument. 
We find analytically a corelation-induced algebraic localization that is robust to a finite strength of the time-reversal-symmetry-breaking parameter,
beyond which all states delocalize. 
This establishes a localization--delocalization transition driven by the interplay between long-ranged correlated hopping and time-reversal symmetry breaking.
In addition to obtaining the static localization phase diagram, we also investigate the dynamical phase diagram through the lens of wavepacket spreading. 
We find that the growth in time of the mean-squared displacement of a wavepacket, which is subdiffusive for the time-reversal symmetric case, becomes diffusive for any finite value of the time-reversal-symmetry-breaking parameter.

\end{abstract}

\maketitle

{\it Introduction.} Anderson localization~\cite{Anderson1958,abrahams201050}, a hallmark phenomenon in condensed matter physics, reveals that  disorder-induced quantum interference leads to exponential localization of wavefunctions in real space, thereby suppressing diffusion. The canonical model of Anderson localization is a tight-binding Hamiltonian with nearest-neighbor hopping and uncorrelated random onsite potentials. In such systems in one dimension (1D), all eigenstates are localized by any infinitesimal amount of disorder~\cite{abrahams1979scaling}.

Despite the apparent robustness of localization in 1D, several notable exceptions exist that have enriched our understanding of localization-delocalization transitions in low-dimensional disordered systems. One prominent example is the Aubry-Andr\'e (AA) model~\cite{AA1980}, where quasiperiodic, and thus correlated, onsite potentials lead to a localization-delocalization transition in 1D. Another class of exceptions, allowing for the presence of extended states even in 1D, is the Power-Law Banded Random Matrix (PLBRM) ensemble~\cite{Levitov1989, Levitov1990, PLRBM, EversMirlin2008} that features uncorrelated hopping amplitudes between two sites decaying as a power law with the distance between them.  The long-range couplings effectively increase the connectivity of the lattice which facilitates transport.

A related actively debated frontier arises in Anderson localization of light~\cite{anderson1985question,segev2013anderson}. 
In three-dimensional disordered media, near-field dipole-dipole interactions between randomly positioned scatterers are believed to hinder a localized phase entirely for vector waves~\cite{skipetrov2014absence,bellando2014cooperative}. Localization is restored only in some specific cases that depend sensitively on the material characteristics of the scatterers~\cite{yamilov2023anderson,yamilov2025anderson}.  
This motivated a model of long-ranged correlated hopping, proposed originally by Logan and Wolynes, and Burin and Maksimov~\cite{logan1985Anderson,logan1987localizability,Burin-Maksimov1989}.
The model combines onsite disorder with translation-invariant hopping correlations mimicking dipole--dipole interactions.
While the correlations, translation invariance, and long-range nature of hopping might be expected to promote delocalization, numerical~\cite{Deng2018Duality} and analytical~\cite{Nosov2019correlation} studies have shown that  the above factors can instead favor   localization. Unlike the uncorrelated hopping of PLBRM, correlations and  translational invariance suppress randomness in hopping pathways, reinforcing localization beyond the convergence of the standard locator expansion~\cite{Anderson1958,Levitov1989,Levitov1990}. Later, several works addressed questions regarding the robustness of correlation-induced localization with respect to the nature of the disorder~\cite{Deng2018Duality, Deng2019Quasicrystals, Kutlin2020_PLE-RG}, partial correlations~\cite{Nosov2019mixtures, Kutlin2021emergent}, dimensionality~\cite{Cantin2018, Deng2022AnisBM} and even interactions~\cite{Santos2016Cooperative} and dynamics~\cite{Deng2024superdiffusion}.

Since Anderson localization is fundamentally rooted in coherent backscattering and quantum interference, one might expect that breaking time-reversal symmetry (TRS) will partially suppress constructive interference between time-reversed paths, or may even destabilize localization, as indeed happens in certain integrable models
\cite{Motamarri2022RDM}. This naturally raises the question of how these expectations reconcile with correlation-induced localization. 
Motivated by this, in this work, we  address the robustness of correlation-induced localization~\cite{Nosov2019correlation} in the presence of TRS breaking.

We show that models with correlated, long-range hopping exhibit robust localization up to a critical strength of the TRS breaking parameter given by the argument of the complex hopping amplitudes. We characterize the resulting localization-delocalization phase diagram by extending the matrix inversion trick of Ref.~\cite{Nosov2019correlation} to TRS-broken systems. Furthermore, we numerically demonstrate that TRS breaking leads to the anomalous dynamics of a wavepacket where the core remains localized but the tails diffuse. This is in contrast to the time-reversal symmetric case, where we observe subdiffusive growth of the wave-packet mean-squared displacement.

\textit{Model.} We consider a one-dimensional disordered model with periodic boundary conditions and fully correlated long-ranged  hopping with complex hopping amplitudes. The Hamiltonian is given by
\begin{align}
H = \sum_n \epsilon_n \ket{n}\bra{n} + \sum_{n \neq m} j_{n-m} \ket{n}\bra{m}, \label{eq: model}
\end{align}
where $\{\ket{n}\}$ denotes the set of real-space basis states and $\epsilon_n\sim\mathcal{N}(0,W)$ are i.i.d., Gaussian random onsite potentials.
The hopping matrix elements,
\begin{align}
j_{n-m} = \frac{J_0 e^{i\theta \text{sign}(n - m)}}{|n - m|^{a}}, \quad (n \ne m), \label{eq: hopping}
\end{align}
are power-law decaying like in the PLBRM, but fully correlated~\cite{foot:TIhopping},
where $J_0$
sets the overall energy scale (without loss of generality we set  $J_0 = 1$), $a > 0$ controls the power-law decay of the hopping amplitude between two sites with the distance between them, and the phase $\theta$ is a tunable parameter that breaks TRS via the difference in hopping phases to the right and left, given by the sign function. The latter is defined as as ${\rm sign}(x) = 1$ for $x>0$ and ${\rm sign}(x) = -1$ for $x<0$. We restrict ourselves to $\theta \in [-\pi/2, \pi/2)$, because the phase diagram is periodic beyond this interval.

For $\theta = 0$, the model reduces to a real, symmetric (time-reversal invariant) hopping matrix , first, suggested for $3$d case in~\cite{logan1985Anderson,logan1987localizability,Burin-Maksimov1989}, which has a localized phase even when the locator expansion diverges at $a<1$~\cite{Deng2018Duality,Nosov2019correlation}, unlike the PLBRM model which has random hopping matrix elements and shows delocalized behavior in the above-mentioned long-range regime.  $\theta \neq 0$  
breaks TRS that induces chiral propagation and suppresses interference between time-reversed paths. 
In the following, we present how TRS breaking (i.e. at $\theta \neq 0$) affects the correlation-induced localization properties.

\begin{figure}[!h]
    \centering
    \includegraphics[width=1.0\linewidth]{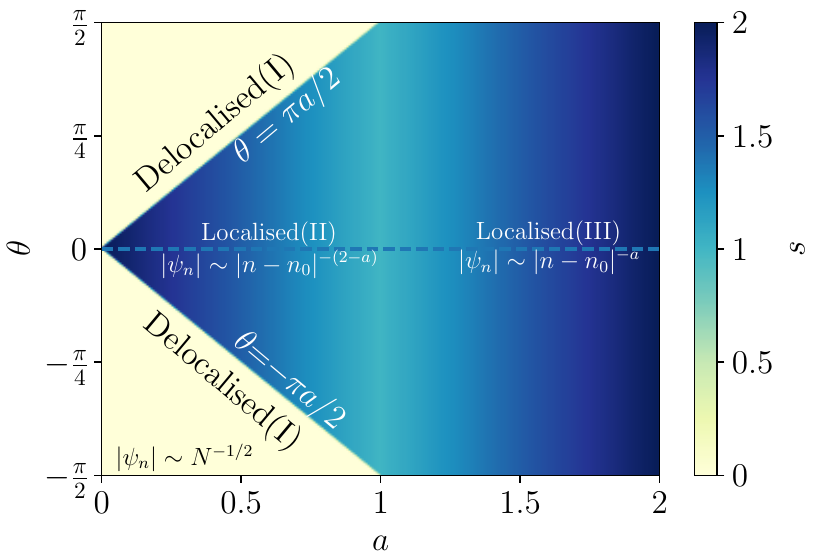}
    \caption{\textit{Phase diagram in the $\theta$–$a$ parameter plane.} The system exhibits a localization-delocalization transition with the delocalized phase (I) appearing for $|\theta| \geqslant \pi a/2$, $a<1$. Further, the localized phase exhibits two flavors [labeled (II) and (III)] characterized by 
    different power-law decays of the eigenstates wavefunctions. The color map indicates the power-law exponent $s$ from Eq.~\eqref{eq:decay-profile} across the parameter space.}
    \label{fig:TRB-Schematic-phase}
\end{figure}

{\it Phase Diagram. } The main result of this work is that the algebraic correlation-induced localization of eigenstates at $a<1$ remains robust even under TRS breaking up to a  critical  parameter value $|\theta|=\theta_c=\pi a/2$, above which TRS breaking destabilizes the localized phase (see Fig.~\ref{fig:TRB-Schematic-phase}). Importantly, the phase boundary, $|\theta_c|=\pi a/2$, being $a$-dependent, leads to a vanishing robustness interval in the integrable limit $a=0$ corresponding to the crossover between the Richardson's and Russian doll models~\cite{Motamarri2022RDM}.

Due to the long-range nature of the model, a vanishing fraction of eigenstates remain plane-wave-like delocalized for $a<3/2$ for weak enough disorder~\cite{Malyshev2000,Malyshev2003,Malyshev2004,Nosov2019correlation} and for {\it any} disorder for $a<1$. However, as in correlation-induced localization~\cite{Nosov2019correlation}, the typical eigenstate decay, $\overline{|\psi(n)|^2_{typ}}\equiv\exp(\overline{\ln|\psi(n)|^2})\sim |n-n_0|^{-2s}$~\footnote{Here and further, the overline denotes an average over disordered Hamiltonians and eigenstates.}, at sites $n$ away from the localization center $n_0$, remains intact and power-law under TRS breaking up to $\theta_c$. The decay exponent $s$ is dual with respect to the PLBRM critical point, $a=1$, i.e., $s(2-a)=s(a)>1$, in direct analogy with correlation-induced localization~\cite{Nosov2019correlation}.  More specifically, typical eigenstates have the following decay profile,
\eq{\overline{|\psi(n)|^2_{typ}}=\begin{cases}
    N^{-1} &\!\!\!\text{ for } |\theta|> \frac{\pi a}2,\, a<1\\
    |n-n_0|^{-2(2-a)} &\!\!\!\text{ for } |\theta|<\frac{\pi a}2, 
 a<1\\
    |n-n_0|^{-2a} &\!\!\!\text{ for }  a>1
\end{cases}.\label{eq:decay-profile}}
We call these three regimes: $(\mathrm{I})$~delocalized phase at strong TRS breaking, $|\theta|> \pi a/2$, and in long-range case, $a<1$; $(\mathrm{II})$~correlation-induced localized phase in the long-range case, $a<1$, with weak TRS breaking, $|\theta|<\pi a/2$; $(\mathrm{III})$~Anderson localized phase in the short-range regime $a>1$. The eigenstate behavior in regimes (I) and (III) is analogous to the one in the long- and short-range phases of the uncorrelated power-law models, like PLBRM or ultrametric matrices~\cite{Fyodorov2009UM}. 

\begin{figure}
    \centering
    \includegraphics[width=\linewidth]{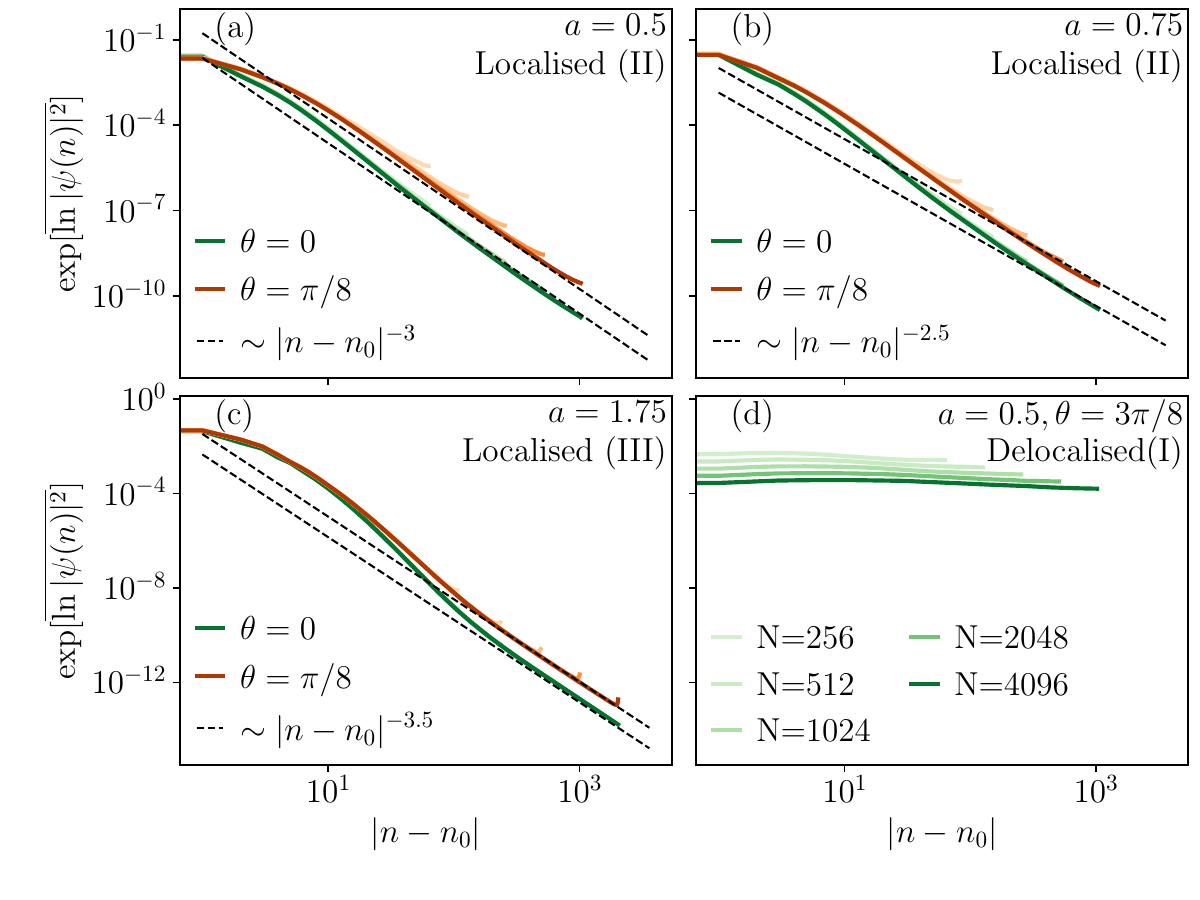}
    \caption{\textit{Spatial decay profile of eigenstates in the bulk of the spectrum.} The eigenstates exhibit algebraic localization, as specified in Eq.~\eqref{eq:decay-profile}, (a-c) which collapses for all system sizes(darker colors represent larger system sizes), and (d) become delocalized for $\theta > \pi a/2$ and $a<1$. Black dashed lines show the analytical predictions of Eq.~\eqref{eq:decay-profile}. The data are averaged over 500 disorder realizations for all system sizes.}
    \label{fig:wave-function-scaling}
\end{figure}

We present numerical evidence for the decay profile of typical eigenstates in Fig.~\ref{fig:wave-function-scaling}. The data show that typical eigenstates follow power-law localization, Eq.~\eqref{eq:decay-profile}, with a power-law exponent that is independent of $\theta$ in both localized regimes (Fig.~\ref{fig:wave-function-scaling}(a)–(c)). In contrast, for $\theta > \pi a/2$ and $a<1$, all eigenstates are extended (Fig.~\ref{fig:wave-function-scaling}(d)). In the following, we derive the whole $\theta-a$ phase diagram, Fig.~\ref{fig:TRB-Schematic-phase} from the hopping spectrum of the above model.

\textit{Spectrum.} Due to the hopping being translation-invariant, the momentum basis is another natural basis for the problem~\cite{foot:TIhopping}. In the Hamiltonian the hopping and the on-site disorder exchange their roles upon the discrete Fourier transformation,

\eq{H=\sum_{p}\tilde{E}_p\ket{p}\bra{p}+\sum_{p,q}\tilde{J}_{p-q}\ket{p}\bra{q},}
where $\ket{p}$ is $p$-momentum basis state, $\tilde{E}_p$ and $\tilde{J}_{p-q}$ are defined as,

\eq{ 
    \tilde{E}_p = \sum_{\substack{n \neq 0}} j_{n} e^{-2\pi i\frac{n p}{N}}, \quad \!\!\!\!
    \tilde{J}_{p-q} = \frac{1}{N}\sum_{m=0}^{N-1} \epsilon_m e^{-2\pi i\frac{(p-q)m}{N}}\label{eq: FT-Ep},}
with integer $0\leq p,q<N$.
The momentum-space hopping terms ${\tilde{J}}_{p-q}$, which come from the Fourier transforms of the real-space onsite potentials $\epsilon_n$, are i.i.d. Gaussian random variables, $\tilde{J}_{p-q}\sim \mathcal{N}(0,N^{-1/2})$. ${\tilde{J}}_{p-q}$ are also correlated in the sense that the hopping amplitude between any two momentum modes separated by a fixed momentum $p-q$ is the same. 

\begin{figure}
    \centering
    \includegraphics[width=\linewidth]{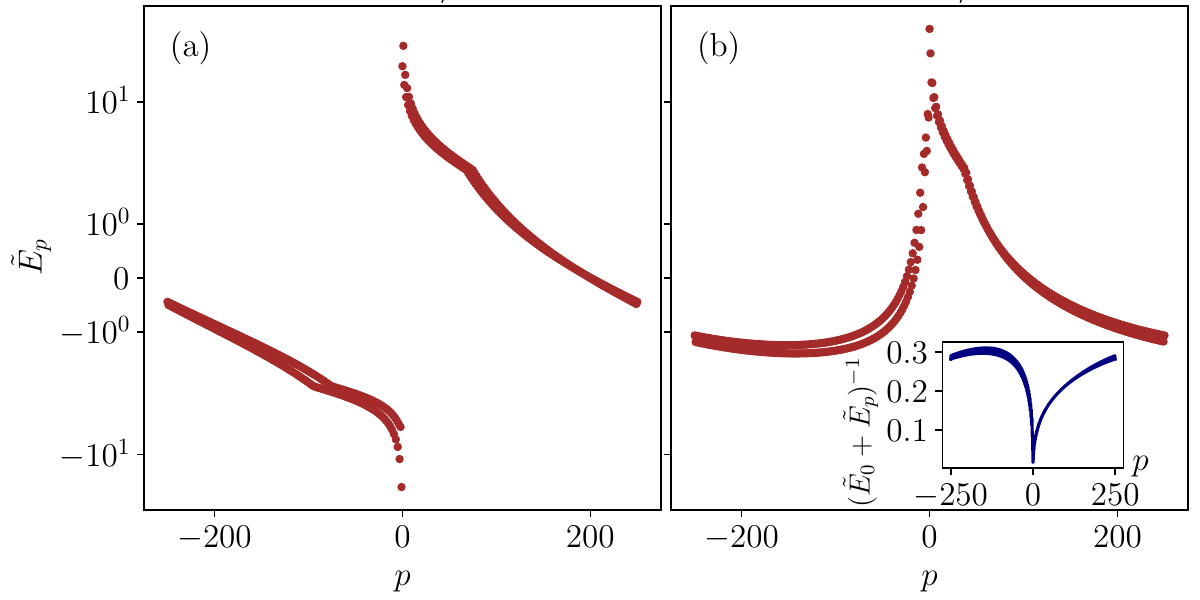}
    \caption{Spectrum of the momentum-space hopping term $\tilde{E}_p$ for $a=0.5$. (a) For $\theta > \pi a/2$, the spectrum diverges in both directions, whereas (b) for $\theta < \pi a/2$, it diverges in only one direction. (Inset) Spectrum of the $M$ matrix, which is bounded from both directions.}
    \label{fig:spectrum-TRB-BM}
\end{figure}

The diagonal part of the momentum-space Hamiltonian, which is nothing but the spectrum of the real-space hopping Hamiltonian,
shows a divergence at small momenta $|p|\ll N$ in the long-ranged regime $a<1$, see Fig.~\ref{fig:spectrum-TRB-BM}. 
For such momenta, $0<|p|\ll N$, which control the long-range structure of eigenstates, the spectrum reads as
\eq{
\tilde{E}_p  = 2\zeta_a +  
          {2\Gamma_{1-a}}\sin\left( \frac{\pi a}{2} + \text{sign}(p) \theta \right) \left( \frac{2\pi |p|}{N} \right)^{a - 1}\!\!\!\!\!\!\!\!\!, 
         \label{eq: Ep+-Ep-}
}
where $\zeta_a$ is the Riemann Zeta function at $a$ and $\Gamma_k$ is the gamma function. Note that, the spectrum diverges in both directions for $|\theta|>\pi a/2$ and diverges only in the positive direction for $|\theta|\leqslant \pi a/2$, see Fig.~\ref{fig:spectrum-TRB-BM}. 
In addition, the spectrum satisfies the relation $\tilde{E}_p(\theta) = -\tilde{E}_p(\theta+\pi)$, which implies that it is sufficient to study the phase diagram in the region $|\theta|\leq\pi/2$.

The high-energy eigenstates, with $|\tilde{E}_p|\gg 1$, remain momentum-space localized (nearly plane waves) under onsite-disorder perturbations because their energy gaps are larger than the momentum-space hopping, i.e., $|\tilde{E}_p-\tilde{E}_q|\gg |\tilde{J}_{p-q}|$ at all $|p|$ or $|q| < p_* \sim N^{1 - 1/(4-2a)}$~\cite{Malyshev2003}. In contrast, for typical states, with $p \sim O(N)$, the perturbation series in momentum space diverges: at any order, the level spacing decreases with $|p|$ faster than the hopping amplitude $|\tilde{J}_{p-q}| \sim N^{-1/2}$. 
At the same time, the real-space perturbation series also diverges, which we discuss next, questioning the origin of a stable algebraic localized phase for $a<1$. Here, the matrix inversion trick (MxIT), which we explain below, helps to understand how correlations in the hopping cause localization of eigenstates which are in the bulk of the spectrum.

{\it Matrix inversion trick and TRS. }The condition for localization is usually understood via Levitov’s resonance counting~\cite{Levitov1989, Levitov1990}, which extends Anderson’s locator expansion argument to the long-range systems. Localization occurs when the hopping amplitude at distance $R$, $|j_R| \sim R^{-a}$, is perturbatively small compared to the mean level spacing, $\delta_{\epsilon_R} = \overline{\min\limits_{m\atop |m-n|<R} \left|\epsilon_n - \epsilon_m\right|}$,  
\begin{equation}
    \frac{|j_R|}{\delta_{\epsilon_R}} < R^{-\epsilon}, \quad \epsilon > 0 \label{eq:Levitov's princ}
\end{equation}
at all $R>\mathcal{O}(1)$.
Here, $\delta_{\epsilon_R} \sim W / R^d$ is the typical level spacing in a ball of radius $R$ for a $d$-dimensional system, and $W$ is the disorder strength variance. For $a > d = 1$, this inequality holds and standard perturbative arguments ensure all eigenstates are localized in all power-law banded models, irrespective of the hopping correlations. 

For $a < 1$, however, $|j_R|/\delta_{\epsilon_R} \sim R^{1-a} \to \infty$, the perturbation series diverges, suggesting delocalization. This reasoning applies to PLBRM models but fails for correlated models: when the series diverges, all perturbative orders contribute, and correlations between matrix elements become essential. The model in \eqref{eq: model} for $\theta=0$ belongs to this category and exhibits localization in this regime due to such correlations. Nosov~\textit{et~al.}~\cite{Nosov2019correlation} resolved this using the \textit{matrix inversion trick} (MxIT), which maps the problem to one with a spectrum bounded on both sides, allowing perturbative expansion to converge even for $a < 1$.

When the spectrum is one-side bounded, in MxIT one uses an offset energy $E_0\sim O(1)$ below the minimum hopping energy, define an inverted matrix,
\eq{M\equiv \left(E_0+j\right)^{-1}=\sum_{p}\frac{1}{E_0+\tilde{E}_p}\ket{p}\bra{p},}
and rewrite the effective hopping term in the spectral bulk as $j_{mn}^{\rm eff}=M_{m-n}(E+E_0 - \epsilon_n)$, keeping the diagonal disorder intact, see, e.g., Eqs.~(15-16) in~\cite{Nosov2019correlation}.

This operation flattens the broad energy spectrum of $j_{m-n}$, making the spectrum of $M$ bounded, as seen in the inset of Fig.~\ref{fig:spectrum-TRB-BM}(b), and allowing for a perturbative treatment in real space. This process enables us to derive an effective hopping, $j_{mn}^{\rm eff}\sim |m-n|^{-(2-a)}$, the details of which are discussed in the Supplemental Material~\ref{app:MIT}. In this effective model, Levitov's criterion, Eq.~\eqref{eq:Levitov's princ}, is now satisfied, extending the localized phase of all the spectral bulk eigenstates to $a<1$ with an effective decay of $2s=2(2-a)$, as given in Eq.~\eqref{eq:decay-profile}.

This MxIT construction applies only when the single-particle spectrum is unbounded in one direction; if it is unbounded in both, the method fails and should be generalized as, e.g., in ~\cite{Motamarri2022RDM}, leading to the ergodically delocalized states. In the latter case, there is no energy $E_0\sim O(1)$ lying outside the spectrum that can be used to define the inverse matrix $M$ without introducing additional divergences, since in the thermodynamic limit the bulk spectrum becomes continuous and gapless.

In our model, the spectrum $\tilde{E}_p$ is unbounded only from one side for $a<1$ and $|\theta|<\pi a/2$, whereas it becomes unbounded in both directions for $|\theta|>\pi a/2$, leading to the breakdown of MxIT. This transition can also be understood from the behavior of the group velocities $v_p=d\tilde{E}_p/dp$, see Fig.~\ref{fig:spectrum-TRB-BM}: for $|\theta|<\pi a/2$, the velocities  have opposite signs for positive and negative $p$, enabling backscattering and promoting localization of the bulk eigenstates. For $|\theta|>\pi a/2$, the velocities share the same sign, leading to the unidirectional motion and, thus, favoring coherent forward scattering and consequently delocalization. Thus, the phase $\theta$ induces a transition for $a <1$ in which the direction(s) of unboundedness directly governs the localization–delocalization physics. By contrast, for $a>1$, the spectrum is bounded for all $\theta$ leading to only a localized phase.

\textit{Wave-Packet Dynamics.} 
We now consider the dynamical phase diagram of the model from the viewpoint of wavepacket dynamics, which shows rich behavior complementing the static phase diagram. To this end, 
we analyze the dynamics of an initially localized wave packet~\cite{Deng2024superdiffusion}. Starting from $\brakett{n}{\psi(0)}=\delta_{n,n_0}$, the spreading is quantified by the generalized $q^{\rm th}$ central moment,

\begin{figure}
    \centering
    \includegraphics[width=\linewidth]{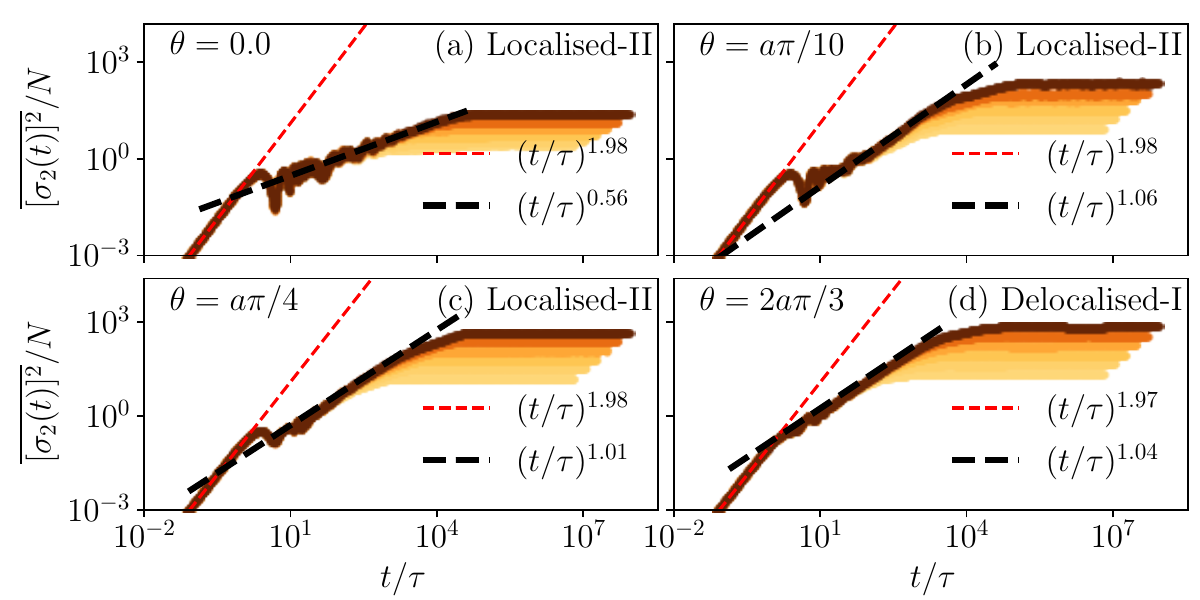}
\caption{\textit{Dynamics of the width of a wave-packet starting from an initially localized state.} The panels show the wave-packet dynamics for $a = 0.25$ at different values of $\theta$, with the darkest color corresponding to $N=8192$: panel \textbf{(a)} (Regime-II with $\theta = 0$) displays an intermediate-time regime with $2\beta = \frac{1}{2-a}\big|_{a=0.25} \approx 0.58$, panels \textbf{(b)} and \textbf{(c)} (Regime-II with $0<\theta<\pi a/2$) exhibit diffusive dynamics between the ballistic front and saturation, and panel \textbf{(d)} ($\theta = 2\pi a/3$) corresponds to the delocalized phase (I), where the intermediate regime is also diffusive. The data is averaged over 500 disorder realizations for $N \leq 1024$ and 100 otherwise.}
    \label{fig:sigma2t-a-0.25}
\end{figure}

\begin{equation}
\sigma_q(t)=\left(\sum_n |n-\langle n\rangle|^q\,|\psi(n,t)|^2\right)^{1/q},
\label{eq:sigmaq_def}
\end{equation}
where $\psi(n,t)=\bra{n}e^{-iHt}\ket{\psi(0)}$ and $\langle n\rangle=\sum_n n\,|\psi(n,t)|^2$. Owing to the power-law-localized eigenstates, different values of $q$ probe distinct spatial regions of the wave packet: small $q$ is sensitive to the core, whereas large $q$ emphasizes the algebraic tails. The transition between these regimes is set by a critical index $q_\ast$. As shown in the Supplemental Material ~\ref{app:sat-sigma-inf},
\begin{equation}
\lim_{t\rightarrow\infty}\left[\sigma_q(t)\right]^q \sim N^{\,q-1/(4-2a)},
\end{equation}
which yields, $q_\ast=\frac{1}{4-2a}$. For $q<q_\ast$, the long-time value of $\sigma_q$  stays finite in the thermodynamic limit, indicating localization of the core. For $q>q_\ast$, the moments retain sensitivity to the extended tails and therefore display nontrivial spreading. This analysis is valid for $a<3/2$ and related to the presence of $p_*\sim N^{1-q_\ast}$ high-energy delocalized states at $a<3/2$ and their contribution to the dynamics.

In the following, we focus on the second moment, $\sigma_2(t)$. Since $q_\ast<2$ for $a<3/2$, $\sigma_2(t)$ is dominated by the tails of the wave packet and therefore exhibits dynamics beyond the trivial ballistic front. The disorder-averaged variance, $\overline{[\sigma_2(t)]^2}$ displays a characteristic three-stage evolution:
\begin{equation}
\overline{[\sigma_2(t)]^2}\propto
\begin{cases}
N^{\Omega}\,(t/\tau)^2 & 0<t<\tau\propto N^{\lambda},\\[4pt]
N^{\Omega}\,(t/\tau)^{2\beta} & \tau<t<t_{\rm sat}\propto N^\delta,\\[4pt]
N^{\gamma} & t>t_{\rm sat},
\end{cases}.
\label{eq:scaling-a-0-0.4-1-1.5}
\end{equation}

\begin{figure}
  \centering

  \begin{minipage}[t]{\columnwidth}
    \centering
    \includegraphics[width=0.75\linewidth]{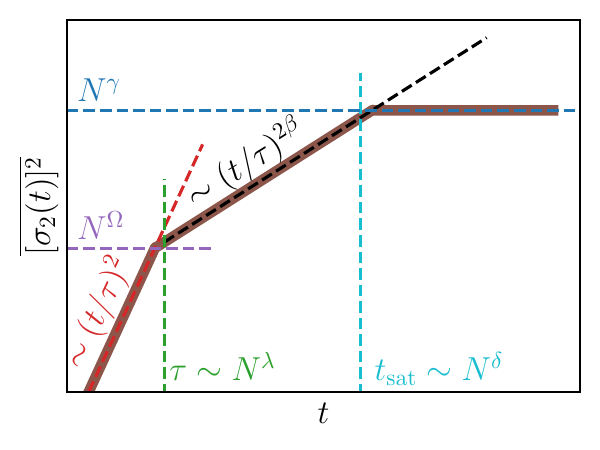}
  \end{minipage}
  \begin{minipage}[t]{\columnwidth}
    \centering
    \scriptsize                  
    \setlength{\tabcolsep}{3pt}  
    \renewcommand{\arraystretch}{1.05}
    \begin{tabular}{|c|c|c|c|c|c|}
      \hline
      & $\Omega$ & $2\beta$ & $\gamma$ & $\lambda$ & $\delta$ \\
      \hline
      (I) $|\theta| > \tfrac{\pi a}{2}$ 
          & \multirow{6}{*}{1} 
          & 1 
          & 2 
          & \multirow{5}{*}{$a-1$}
          & $a$ \\
      \cline{1-1} \cline{3-3} \cline{4-4} \cline{6-6}
      (II) $\theta=0$, $a \leq 1$ 
          &  
          & $\tfrac{1}{2-a}$ 
          & \multirow{4}{*}{$1-\tfrac{1}{4-2a}$}
          & 
          & $\tfrac{1}{2}$ \\
      \cline{1-1} \cline{3-3} \cline{6-6}
      (II) $\theta=0$, $a>1$ 
          &
          & \multirow{3}{*}{1} 
          & 
          &
          & \multirow{3}{*}{$a-\tfrac{1}{4-2a}$} \\
      \cline{1-1}
      (II) $0<|\theta|<\frac{\pi a}{2}$ 
          &
          &
          &
          &
          &  \\
      \cline{1-1}
      (III) $1<a<\tfrac{3}{2}$ 
          &
          &
          &
          &
          &  \\
      \hline
      (III) $a>\tfrac{3}{2}$ 
          & 0 & -- & 0 & 0 & 0 \\
      \hline
    \end{tabular}
  \end{minipage}
\caption{\textit{Schematic illustration of the dynamical windows (top) and their associated scaling exponents (bottom)}. The top panel shows a schematic diagram of the different regimes in the dynamics of the wave-packet width: the red, black, and blue dashed lines represent the ballistic, intermediate (sub-)diffusive, and saturation regimes, respectively, while the green and blue vertical dashed lines mark the crossover timescales $\tau \sim N^{\lambda}$ and $t_{\rm sat} \sim N^{\delta}$ between these windows. The table reports the corresponding exponents $\Omega$, $2\beta$, $\gamma$, $\lambda$, and $\delta$ governing the dynamics
in different windows.}

  \label{fig:schematic-and-table}
\end{figure}

We summarize the exponents ($\Omega,2\beta,\gamma,\lambda,\delta$) for the different parameter windows in Table 1 of Fig. \ref{fig:schematic-and-table}, and present numerical evidence and analytical understanding of the ballistic decay in the Supplementary Material, \ref{app:ballistic-scaling} and\ref{app:Dyn-for-a-theta}, respectively. We also discuss the origin of the short-time ballistic decay and the scaling of the long-time width, i.e., $\gamma$, there.

Remarkably, although the eigenstate properties in the bulk of the spectrum yield three distinct regimes, Fig.~\ref{fig:TRB-Schematic-phase}, the dynamics further splits. For example, in the correlation-induced localized phase (II), the eigenstate  spatial decay is insensitive to the choice of $\theta$. In contrast, the dynamics being sensitive to relative phases and correlations between eigenstates displays qualitatively different behavior: in the intermediate time window, second moment is diffusive for $\theta\ne 0$, while it becomes subdiffusive for $\theta=0$, see Fig.~\ref{fig:sigma2t-a-0.25}. The (sub)diffusive behavior originates from an extensive yet asymptotically vanishing fraction ($p_*/N\sim N^{-q_\ast})$ of high-energy delocalized states for $0<a<3/2$~\cite{Malyshev2003,Malyshev2005}, which dominate the dynamics for lower moments $q<q_\ast$.

\textit{Conclusion and Outlook.}
We summarize the main findings of this work. We investigated localization phenomena in long-range correlated hopping models with broken time-reversal symmetry and established two key results. 
First, on the static side, we showed that the power-law–localized eigenstates in the bulk of the spectrum remain robust against time-reversal symmetry breaking up to $|\theta|=\pi a/2$, beyond which all states become delocalized. The resulting phase diagram was explained analytically by extending the matrix inversion trick to finite $\theta$, clarifying the stability of the correlation-induced localized phase.

Second, from the dynamical perspective, we analytically determined the saturation properties of the generalized moments $\sigma_q(t)$, demonstrating how different values of $q$ selectively probe the core or the tails of the wave packet. We further showed numerically that the experimentally accessible variance $\overline{[\sigma_2(t)]^2}$ is strongly sensitive to time-reversal symmetry breaking unlike the static eigenstate  spatial decay and exhibits distinct diffusive or subdiffusive behavior depending on $\theta$.

While we have established a clear understanding of both the early-time ballistic regime and the saturation scale of the wave packet, a full analytical description of the (sub)diffusive intermediate-time scaling remains open. This regime, which also appears in light-localization numerics~\cite{sgrignuoli2020subdiffusive,sgrignuoli2022subdiffusive}
involving dipole-dipole interactions which is basically long-range correlated hopping, offers an exciting opportunity for further exploration. We anticipate that a deeper analytical and experimental investigation of this connection will provide valuable insight into anomalous transport in generic long-range correlated systems.

An independent extension concerns the robustness of correlation-induced localization to non-Hermiticity. Although the present model allows complex hopping phases, the Hamiltonian remains Hermitian; introducing asymmetric non-hermitian hopping would generalize the Hatano--Nelson model~\cite{hatano1996localization} to long-range correlated systems. Whether the algebraically localized phases identified here persist under non-Hermitian spectral flow and associated skin effects remains an open question.

\begin{acknowledgements}
\textit{Acknowledgments.}
I.M.K. acknowledges support by the European Research Council under the European Union’s Seventh Framework Program Synergy ERC-2018-SyG HERO-810451. B.P. acknowledges support from NORDITA through the NORDITA Visiting Ph.D. Fellowship Program, as part of which this work was carried out. 
B.P. and S.R. acknowledge support from the Department of Atomic Energy, Government of India, under Project Nos. RTI4013 and RTI4019. 
S.R. acknowledges support from SERB-DST, Government of India, under Grant No. SRG/2023/000858, and from a Max Planck Partner Group grant between ICTS-TIFR, Bengaluru and MPIPKS, Dresden.
J.H.B received funding from the European Research Council (ERC) under the European Union’s Horizon 2020 research and innovation program (Grant Agreement No. 101001902) and the Knut and Alice Wallenberg Foundation (KAW) via the project Dynamic Quantum Matter (GrantNo. 2019.0068).
\end{acknowledgements}

\bibliography{Lib}


\newpage

\setcounter{equation}{0}
\setcounter{figure}{0}
\setcounter{page}{1}
\renewcommand{\theequation}{S\arabic{equation}}
\renewcommand{\thefigure}{S\arabic{figure}}
\renewcommand{\thesection}{S\arabic{section}}
\renewcommand{\thepage}{S\arabic{page}}

\onecolumngrid

\begin{center}
    {\bf Supplementary Material: Robust Correlation-Induced Localization Under Time-Reversal Symmetry Breaking}\\
    Bikram Pain, Sthitadhi Roy, Jens H. Bardarson, Ivan M. Khaymovich\\
    
\end{center}

In this supplementary material, we discuss the Matrix Inversion Trick in the context of our model with broken time-reversal symmetry. Furthermore, we analytically derive the scaling of the saturation of the width of the wave packet, its ballistic scaling at short times \(t\), and present numerical evidence of different exponents characterizing the dynamics for several values of \(\theta\) and \(a\).

\section{Matrix inversion trick (MxIT)}
\label{app:MIT}

This section provides a concise derivation of the matrix inversion trick (MxIT) used in the main text to demonstrate localization of spectral-bulk states in long-range Burin-Maksimov-type models. We begin the derivation for $\theta = 0.0$, as done in~\cite{Nosov2019correlation}, and briefly discuss its impact on time-reversal symmetry breaking.
\subsection{MxIT Construction for One–Sided Unbounded Spectrum}

For $a<1$, the momentum–space hopping spectrum diverges at small momenta,
\begin{equation}\label{Supp:E_p}
    \tilde{E}_p \sim \left(\frac{N}{p}\right)^{1-a}, \qquad |p|\ll N,
\end{equation}
and the energy spacing between nearby momentum states becomes parametrically smaller than the disorder-induced matrix elements. As a result, perturbation theory in either momentum or real space fails. MxIT resolves this by inverting the hopping operator after shifting by a constant. Starting from
\begin{equation}
    (E - \epsilon_n)\psi_E(n) = \sum_{m\neq n} j_{n-m}\,\psi_E(m)\,\text{ or } \quad     J\ket{\psi_E} = (E - \epsilon)\ket{\psi_E},
\end{equation}

with $J$ being the hopping  operator, $\ket{\psi_E}$ the eigenstate, and $\psi_E(n)=\braket{n|\psi_E}$, we define the inverse operator
\begin{equation}
    M \equiv (J + E_0)^{-1} = \sum_{p} \frac{1}{\tilde{E}_p + E_0}\ket{p}\bra{p}, \label{eq:MIT-def}
\end{equation}
where $\ket{n}$ and $\ket{p}$ are the $n$-site and $p$-momentum basis states, respectively, $E_0\sim O(N^0)$ is chosen such that $-E_0$ lies outside the original spectrum in the thermodynamic limit. This is possible only when the latter is unbounded in one direction, i.e., $|\theta|<\pi a/2$, as discussed in the main text. $M$ is diagonal in momentum space, and therefore translationally invariant in real space. Multiplying by $M$ yields a effective Schr\"odinger equation,
\begin{equation}
    (E_{\mathrm{eff}}-\epsilon_n)\psi_n =
    \sum_{m\neq n} j_{nm}^{\rm eff}\psi_m,
\end{equation}
with
\begin{equation}
    E_{\mathrm{eff}} = E + E_0 - \frac{1}{M_0}, \qquad
    j_{n-m}^{\rm eff} = \frac{M_{n-m}}{M_0}(E+E_0-\epsilon_m),
\end{equation}
where $M_0=M_{n-n}$. For $a<1$, the inverted spectrum behaves as $\tilde{M}_p=(\tilde{E}_p + E_0)^{-1}\sim (N/p)^{a-1}$, and,thus, its Fourier transform gives $M_{n-m} \sim |n-m|^{-(2-a)}$ cf. Eq.~\eqref{Supp:E_p}. Thus the effective hopping becomes
    $j_{n-m}^{\rm eff} \sim |n-m|^{-(2-a)},$
allowing Levitov's criterion, Eq.~\eqref{eq:Levitov's princ}, to be satisfied,
\begin{equation}
    \frac{|j_{n,n+R}^{\rm eff}|}{\delta_{\epsilon_R}} \sim R^{-(1-a)} \to 0,
\end{equation}
and ensuring localization of the bulk eigenstates for all $a<1$. This yields the spatial decay exponent quoted in the main text in Eq.~\eqref{eq:decay-profile}.

\subsection{Breakdown for $|\theta|\ge\pi a/2$}

When $|\theta| \ge \pi a/2$, the spectrum $\tilde{E}_p$ diverges in both directions. In this case no finite shift $E_0$ exists that places $-E_0$ outside the entire spectral support, making $(J+E_0)^{-1}$ ill–defined. Consequently, the MxIT mapping cannot be constructed and no rapidly decaying effective hopping emerges, leading to delocalization of the corresponding states.

In summary, MxIT replaces a divergent momentum–space description with a real–space model having effective hopping $|n-m|^{-(2-a)}$, restoring perturbative control and stabilizing localization for $a<1$ when the spectrum is one–sided unbounded. Time–reversal symmetry breaking changes the topology of the spectrum, determining whether the MxIT and therefore localization persists.

\section{Saturation of $\lim_{t\rightarrow\infty}[\sigma_q(t)]^q$}
\label{app:sat-sigma-inf}
\begin{figure}[!b]
    \centering
    \includegraphics[width=0.9\linewidth]{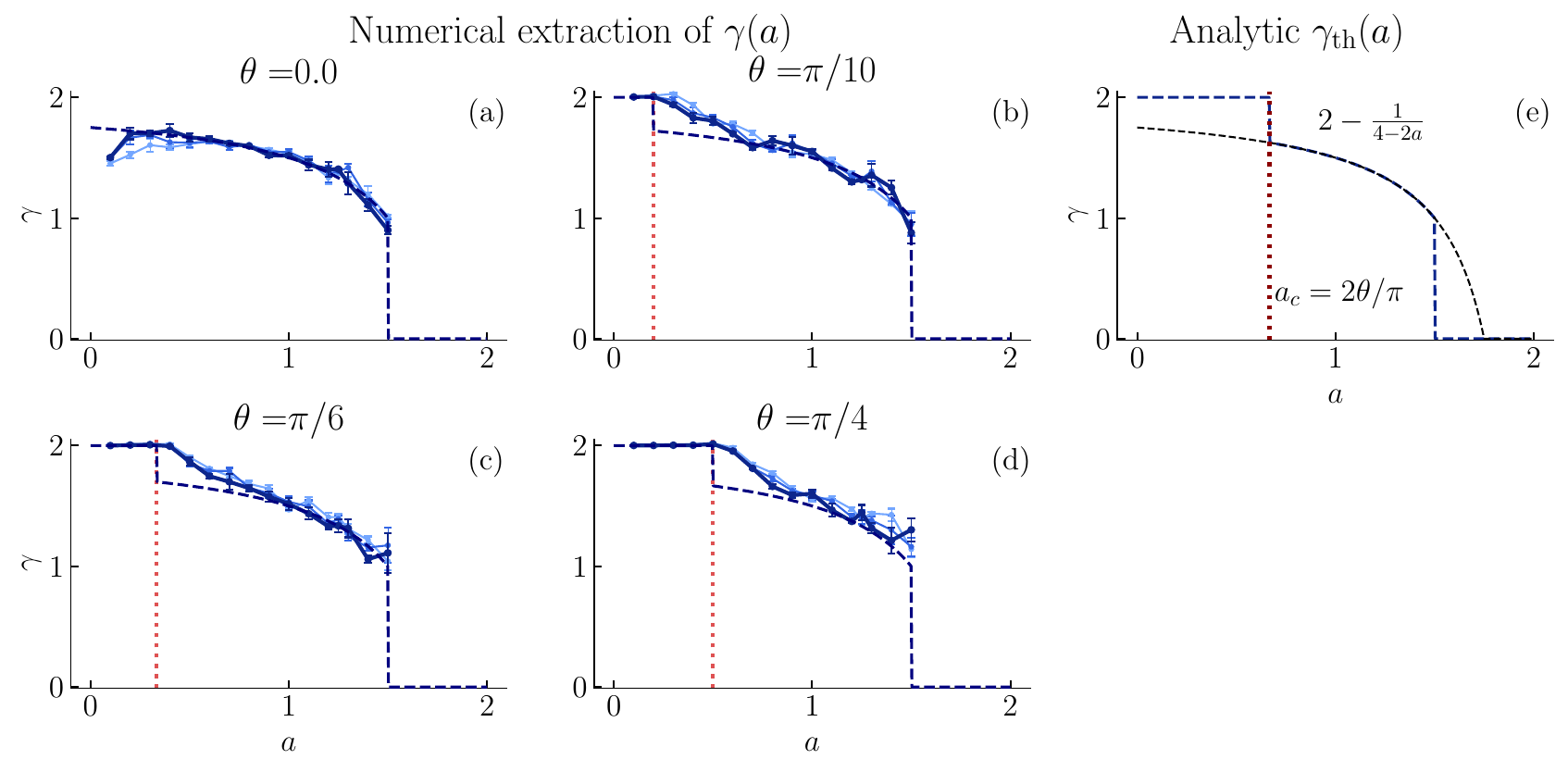}
\caption{\textit{Scaling exponent $\gamma(a)$ extracted from the late--time saturated width $\overline{[\sigma_2(t\to\infty)]^2}\sim N^{\gamma}$ for different values of $\theta$.} Panels (a)--(d) show the finite size estimates of $\gamma(a)$ obtained from sliding-window fits over increasing system sizes $N$ from $[256-2048]$ to $[1024-8192]$, indicated by progressively darker blue tones. The dark blue dashed line denotes the asymptotic prediction of Eq.~\ref{eq: gamma-asym},
and the red vertical line marks the phase transition point
$a_c = 2\theta/\pi$.
Panel (e) displays the corresponding analytic prediction:
$\gamma = 2$ for $a < a_c$, 
$\gamma=2-1/(4-2a)$ for $a_c < a < 3/2$ (by black dashed line), 
and $\gamma = 0$ for $a > 3/2$.
The excellent agreement between numerics and theory demonstrates that the crossover position is controlled by $\theta$, and that the asymptotic scaling is reached systematically upon increasing $N$.
}
\label{fig:gamma-a}
\end{figure}

In this subsection, we discuss the generalized central moments of the wave packet at large time $t$. First, we have considered the $\theta=0.0$ case, which represents time-reversal symmetry, as verified numerically in Fig~\ref{fig:gamma-a}. For $\theta \neq 0$, we have argued that strong finite-size effects are present, and the behavior should be the same as in the $\theta=0.0$ case in the thermodynamic limit. 

In our model in \eqref{eq: model}, there exist both power-law localized states and delocalized states. We approximate them as
\begin{equation}
|\psi_m(n)|^2=\begin{cases}
    \frac{\Gamma_{s}}{\sqrt{\pi}\,\Gamma_{{s}-1/2}}\dfrac{1}{\left[(m-n)^2+1\right]^{s}} & \text{for localized eigenstates},\\[6pt]
    \dfrac{1}{N} & \text{for delocalized eigenstates},
\end{cases}
\end{equation}
where the number of delocalized states scales as $p_*\sim N^{1-q_\ast}$ with $q_\ast=1/(4-2a)$ for $a<3/2$~\cite{Malyshev2003,Malyshev2005}, while all states are power-law localized for $a>3/2$. Here and further, without loss of generality, we approximate the localization centers $m$ to be evenly distributed over the sample.
In the long-time limit, in the absence of spectral degeneracies, the exponent $e^{-i (E_m-E_{m'}) t}$ reduces to the Kronecker delta $\delta_{m,m'}$, leading to
\begin{equation}\label{eq:sigma_q_sat_int}
\overline{[\sigma_q(t\to\infty)]^q}
= \sum_{m} |\psi_m(0)|^2 \sum_n n^q |\psi_m(n)|^2
\simeq N^{q-q_\ast} + \sum_{m} |\psi_m(0)|^2 \int_{-N/2}^{N/2}  \frac{\Gamma_{s}}{\sqrt{\pi}\,\Gamma_{{s}-1/2}}\frac{(m-x)^q\,dx}{\left(x^2+1\right)^{s}} .
\end{equation}

Here, we have split the sum into contributions from the $N^{1-q_\ast}$ delocalized states and the remaining localized states. The integral over $x = m-n$ in Eq.~\eqref{eq:sigma_q_sat_int} can be analyzed by simple power counting: for $2s>q+1$ it converges at large $|x| \sim N$, whereas for $2s<q+1$ it yields a contribution $N^{1+q-2s}\gg 1$. The sum over $m$ contains a polynomial of degree $q$ in $m$, giving the same scaling. For example, considering $q=2$, the integral yields
\begin{equation}\label{eq:sigma_2_sat}
\overline{[\sigma_2(t\to\infty)]^2}\simeq N^{2-q_\ast} + 2\left(\frac{1}{2s-3}+A\,N^{3-2s}\right),
\end{equation}
confirming the scaling above, with a $s$-dependent constant $A$. Therefore, for general $q$, the saturation scales as
\begin{equation}\label{eq:sigma_q_sat_res}
\overline{[\sigma_q(t\to\infty)]^q} \simeq N^{q-q_\ast} + O(1) + N^{(1+q)-2s}.
\end{equation}

The effective decay exponent is $s=\max(2-a,a)$. This results in the first term dominating for $a<3/2$, irrespective of $q$. Therefore, for a generic $q$, we have:

\begin{equation}\label{eq:sigma_q_sat_BM}
\overline{[\sigma_q(t\to\infty)]^q} \simeq 
\left\{
\begin{array}{ll}
N^{q-1/(4-2a)} &  a<\frac{3}{2} \\
N^{(1+q)-2a} &  \frac{3}{2}<a<\frac{1+q}{2} \text{ and } q>2 \\
O(1)  &  a>\frac{3}{2}, \frac{1+q}{2} \\
\end{array}
\right.
\end{equation}

This indicates that the saturation value is $N$-independent only when $a > (q+1)/2$ and $a > 3/2$.
 In the delocalized regime (I), (i.e. $|\theta|\geqslant \pi a/2$ with $a<1$) as all eigenstates are delocalized (i.e., $q_*=0$), only the first term contributes, which gives $\left[\sigma_q(t \to \infty)\right]^q \simeq N^{q}$. In the localized regimes (II) and (III) (untill $a < 3/2$), a strong finite-size effect is observed due to the difference in the prefactor in Eq. \ref{eq: Ep+-Ep-} for $|\theta| > 0$. This difference alters the effective power-law scaling of the two parts of the spectrum, $\ln E_{p_+}/\ln N$ and $\ln E_{p_-}/\ln N$. These two become of the same order only at $N \sim O(10^{100})$ due to the slow power-law decay of the eigenstates. In the thermodynamic limit, we anticipate the saturation value to be independent of $\theta$, and for $q=2$ only the second term in \eqref{eq:sigma_q_sat_BM} will dominate. Finite-size effects also exist for $a > 3/2$ in the localized regime (III), as the saturation to a $O(1)$ number decays as $N^{3-2a}$, which is quite small for $3/2 < a < 2$. So, in the thermodynamic limit we expect that  $\overline{[\sigma_2(t)]^2}\propto N^{\gamma}$, with  

\eq{\gamma=\begin{cases}
    2 & \text{ for } |\theta|\geqslant \pi a/2,\\
    2-1/(4-2a)& \text{ for } |\theta|<\pi a/2, a<3/2\\
    0 & \text{ for } a\geq 3/2
\end{cases}\label{eq: gamma-asym}}

\section{Ballistic scaling of $[\sigma_2(t)]^2$}\label{app:ballistic-scaling}

We analyze the short-time spreading of the wave packet by evaluating the second moment of the position operator perturbatively. Expanding
\begin{equation}
\braket{n^q(t)}=\braket{\psi(0)|e^{iHt}n^qe^{-iHt}|\psi(0)}
\end{equation}
to second order in time gives
\begin{equation}
\braket{n^q(t)}=
\braket{n^q(0)}
+ it\,\braket{[H,n^q]}
- \frac{t^2}{2}\braket{[H,[H,n^q]]}
+ \mathcal{O}(t^3),
\end{equation}
where $\braket{\cdot}\equiv\braket{\psi(0)|\,\cdot\,|\psi(0)}$. For a localized initial state $\ket{\psi(0)}=\ket{n=0}$, the first two terms vanish, so the leading contribution is quadratic in time,
\begin{equation}
\braket{n^q(t)} \simeq t^2 \braket{H\,n^q\,H}.
\label{eq:short}
\end{equation}

Expressing the Hamiltonian in its eigenbasis,
\begin{equation}
H=\sum_{m}E_m|m\rangle\langle m|,\qquad
|m\rangle=\sum_n C^m_n |n\rangle,
\end{equation}
Eq.~\eqref{eq:short} becomes
\begin{equation}
\braket{Hn^qH} =
\sum_{m,m^\prime} E_m E_{m^\prime}\,
C_{0}^{m\ast}C_{0}^{m^\prime}
\sum_n n^q C_n^m C_n^{m^\prime\ast}.
\label{eq:double}
\end{equation}

The scaling of \eqref{eq:double} depends on the spectral composition. For $a<3/2$, we decompose the sum as
\begin{equation}
\sum_{m,m^\prime} =
\sum_{\rm deloc,deloc} +
\sum_{\rm loc,deloc} +
\sum_{\rm loc,loc}.
\end{equation}

\paragraph{Delocalized–delocalized contribution.}
For delocalized eigenstates, $C_n^m\simeq N^{-1/2}e^{i2\pi np_m/N}$ and $E_m\sim (p_m/N)^{a-1}$, so
\begin{equation}
\sum_{m,m^\prime \in \mathrm{deloc}}
E_mE_{m^\prime}C_{0}^{m\ast}C_{0}^{m^\prime}
\sum_n n^q C_n^m C_n^{m^\prime\ast}
\sim \sum_n n^q \left|\frac{1}{N}
\sum_{p < p_\ast}
(p/N)^{a-1} e^{i2\pi np/N}\right|^2.
\end{equation}
Converting the sum to an integral gives
\begin{equation}
I(N,n)=\frac{1}{N}\int_{0}^{p_\ast} (p/N)^{a-1} e^{i2\pi n p/N}\,dp.
\end{equation}
For $n \gg N/p_{\ast}$, the integral oscillates rapidly for large $|p|$, and the dominant contribution arises from the lower limit. Therefore, we may extend the upper limit to $\infty$ without significant error, yielding the inverse Fourier transform result
\begin{equation}
I(N,n) \sim n^{-a}.
\end{equation}
Thus,
\begin{equation}
\sum_{\rm deloc,deloc} \sim
\sum_n n^2 |I(N,n)|^2
\sim \sum_n n^{2-2a}
\sim N^{3-2a}.
\end{equation}

\paragraph{Localized–delocalized contribution.}
When one eigenstate is localized and the other is delocalized, Eq.~\eqref{eq:double} becomes
\begin{equation}
\sum_{m\in {\rm deloc},\, m^\prime\in {\rm loc}} 
E_m E_{m^\prime}\,
C_{0}^{m\ast}C_{0}^{m^\prime}
\sum_n n^q C_n^m C_n^{m^\prime\ast}
\approx
\sum_{m^\prime\in{\rm loc}} E_{m^\prime} C_{0}^{m^\prime}
\sum_{n} n^q C_n^{m^\prime\ast}
\bigg(\frac{1}{N}\sum_{p=0}^{p_\ast}E_p e^{i2\pi n p/N}\bigg),
\end{equation}
where for delocalized states we have approximated 
$C_n^m=N^{-1/2}e^{i2\pi np/N}, C_0^{m}\sim N^{-1/2} $. The momentum sum is dominated by small $p$, and extending the upper limit to $\infty$ gives the inverse Fourier transform of the spectrum,
\eq{\frac{1}{N}\sum_{p=0}^{\infty} E_p e^{i2\pi np/N}\sim |n|^{-a}.}

For a localized state,$
|C_n^{m^\prime}| \sim |n-x_{m^\prime}|^{-s},$ with $s = \max(2-a,a),$
hence
\eq{
\sum_{n} n^2 C_n^{m^\prime\ast} |n|^{-a}
\sim \sum_{n} n^{2-a-s}
\sim N^{3-a-s}.}

Since only $\mathcal{O}(1)$ localized states have appreciable overlap with $|0\rangle$, and $E_{m^\prime}=\mathcal{O}(1)$,
\begin{equation}
\sum_{m\in {\rm deloc},\, m^\prime\in {\rm loc}} 
E_m E_{m^\prime}\,
C_{0}^{m\ast}C_{0}^{m^\prime}
\sum_n n^q C_n^m C_n^{m^\prime\ast}
\propto
\begin{cases}
N^{-1}, & a<1,\\[4pt]
N^{1-2a}, & 1<a<3/2,
\end{cases}
\label{eq:locdeloc}
\end{equation}
which is subleading compared to the deloc–deloc contribution $\sim N^{3-2a}$ for all $a<3/2$.

\paragraph{Localized–localized contribution.}
For purely localized eigenstates, the overlap
\[
\sum_x x^q C_x^m C_x^{m^\prime\ast}
\sim \int dx\, x^q |x-x_m|^{-s}|x-x_{m^\prime}|^{-s}
\]
converges for $0<a<2$ with $q=2$ and is $O(1)$ in system size. Moreover, $C_0^m$ is appreciable only for eigenstates centered within a localization length of the origin, so only $O(1)$ localized states contribute appreciably. In this sector, the energy level spacing satisfies $\delta E\sim 1/N$ and $E_m\sim O(1)$, giving
\[
\sum_{\rm loc,loc} E_m E_{m^\prime} C_0^{m\ast} C_0^{m^\prime}
\sim N^0.
\]
Thus the loc–loc sector does not produce any $N$-dependent enhancement.

\paragraph{Summary.}
Collecting all scaling contributions for the second moment, we obtain
\begin{equation}
[\sigma_2(t)]^2
\propto
\begin{cases}
N\bigg(\frac{t}{N^{a-1}}\bigg)^2, & 0<a<3/2,\\
t^2, & a>3/2,
\end{cases}
\label{eq:final_sigma}
\end{equation}
establishing that the initial wave-packet spreading is ballistic ($\sigma_2(t)\sim t$), but exhibits a strong system-size–dependent prefactor which disappears once delocalized states vanish at $a\geqslant3/2$. This supports the scaling of $\Omega$ and $\tau$ in the Fig.~\ref{fig:schematic-and-table}.

\section{Dynamics for different $a$ and $\theta$}\label{app:Dyn-for-a-theta}

In this section, we provide additional numerical evidence supporting the scaling behavior of the wave-packet width and the associated dynamical exponents summarized in Fig.~\ref{fig:schematic-and-table} of the main text.

\begin{figure}
    \centering
    \includegraphics[width=0.8\linewidth]{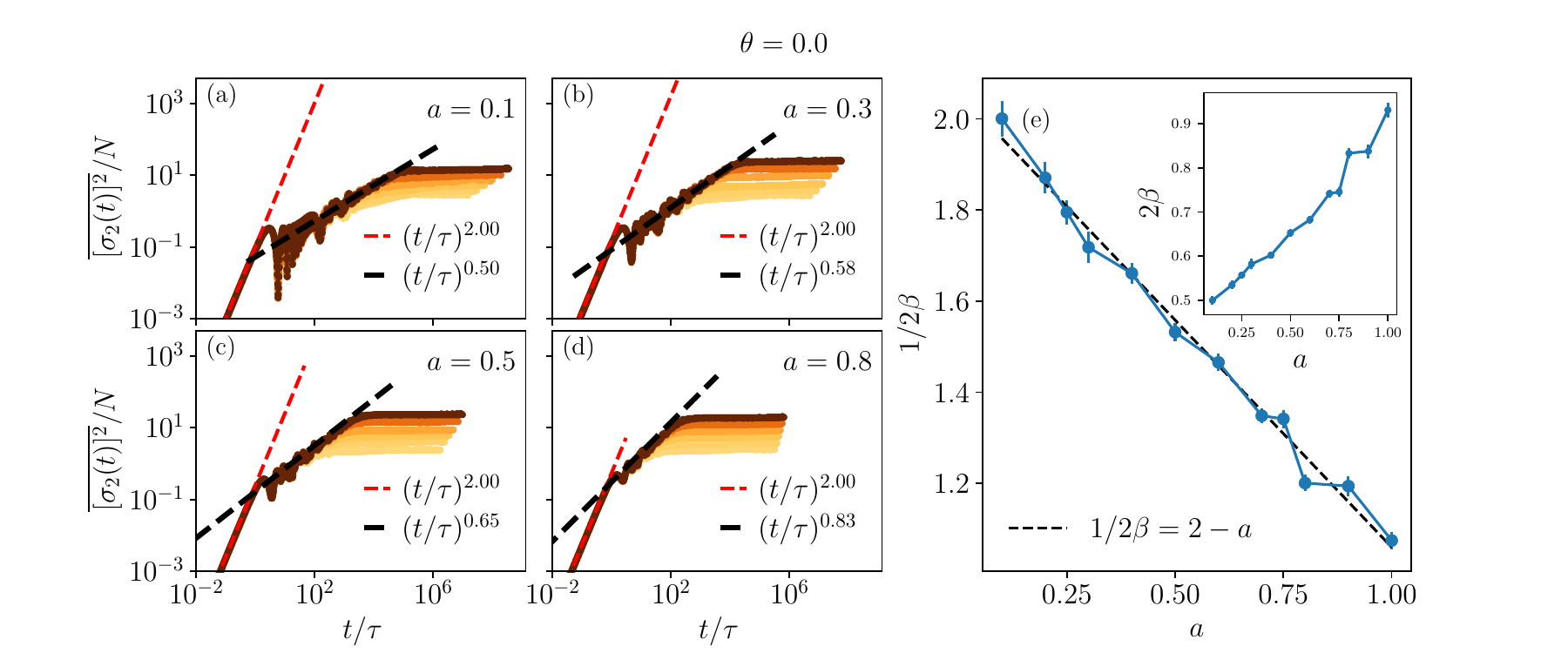}
\caption{(Left) Dynamics of the wave packet along the time-reversal symmetry (TRS)-preserving line, $\theta = 0$. (Right) Scaling of the dynamical exponent, $1/2\beta$ as a function of $a$. (Right inset) The raw data of $2\beta$ versus $a$. Darkest color represents the largest system size, i.e., $N=8192$, and the data is averaged over 500 realizations for $N \leq 1024$ and 100 otherwise.}
    \label{fig:theta-0-2beta-a}
\end{figure}

We first consider the time-reversal-symmetric case, 
shown in Fig.~\ref{fig:theta-0-2beta-a}. Along the $\theta=0$ line, the width of the wave packet exhibits sub-diffusive growth over an extended intermediate-time window (i.e. $O(N^{a-1})<t<\sqrt{N}$). The corresponding sub-diffusive exponent follows the scaling
\eq{2\beta \simeq \frac{1}{2-a} \quad \text{for } a<1,}
while for $a>1$ the dynamics crosses over to diffusion with $2\beta=1$.

\begin{figure}
    \centering
    \includegraphics[width=0.5\linewidth]{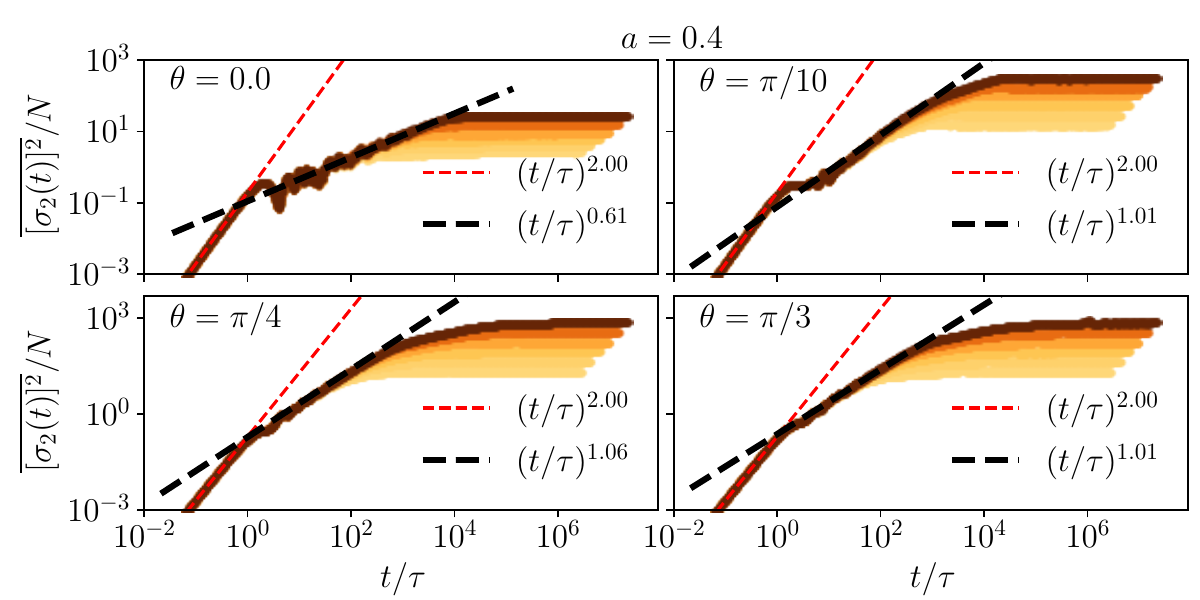}
\caption{\textit{Dynamics of the wave-packet width for $a=0.4$ and different values of $\theta$.} Diffusive scaling is observed for $|\theta|>0$. Darkest color represents the largest system size, i.e., $N=8192$, and the data is averaged over 500 realizations for $N \leq 1024$ and 100 otherwise.}
    \label{fig:a-0.4}
\end{figure}

Next, we show the dynamics along fixed-$a$ cuts for different values of the TRS-breaking parameter $\theta$. For $a=0.4$, shown in Fig.~\ref{fig:a-0.4}, the wave-packet width exhibits diffusive growth for any nonzero $|\theta|$, indicating that even an infinitesimal breaking of TRS destabilizes the sub-diffusive regime present at $\theta=0$.

\begin{figure}
    \centering
    \includegraphics[width=0.5\linewidth]{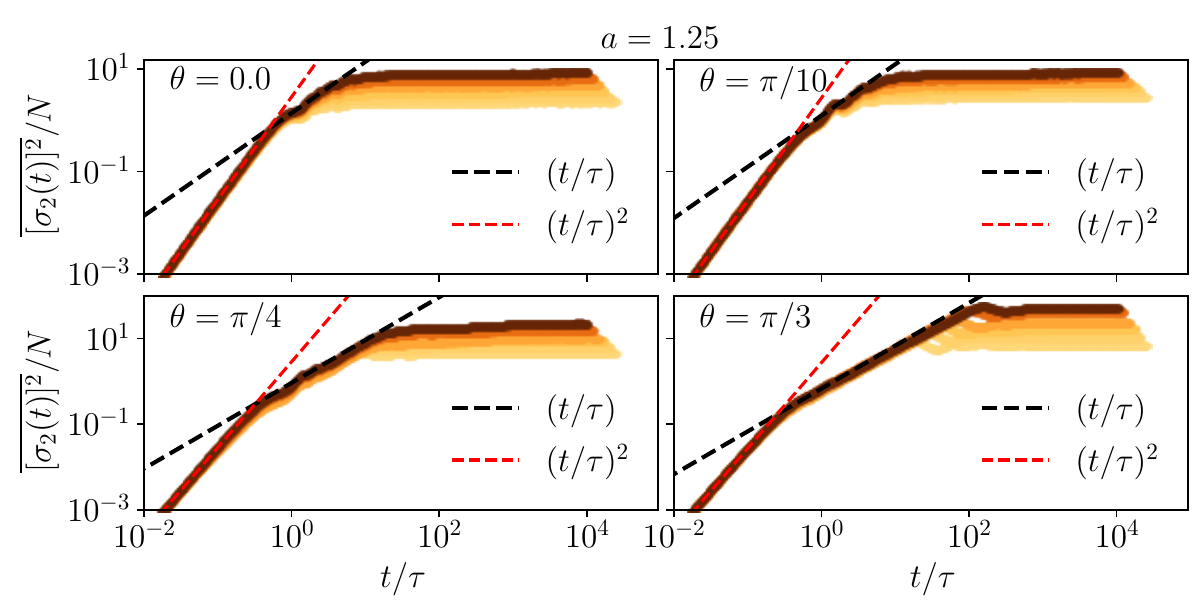}
\caption{\textit{Dynamics of the wave-packet width for $a=1.25$ and different values of $\theta$.} The diffusive regime widens with increasing $|\theta|$. Darkest color represents the largest system size, i.e., $N=8192$, and the data is averaged over 500 realizations for $N \leq 1024$ and 100 otherwise.}
    \label{fig:a-1.25}
\end{figure}

In Fig.~\ref{fig:a-1.25}, corresponding to $a=1.25$, the dynamics remains diffusive for all values of $\theta$. However, the extent of the diffusive time window increases with increasing $|\theta|$. This behavior reflects the fact that stronger TRS breaking induces increasingly random phases in the eigenstates, thereby sustaining diffusion over longer times.

\begin{figure}
    \centering
    \includegraphics[width=0.5\linewidth]{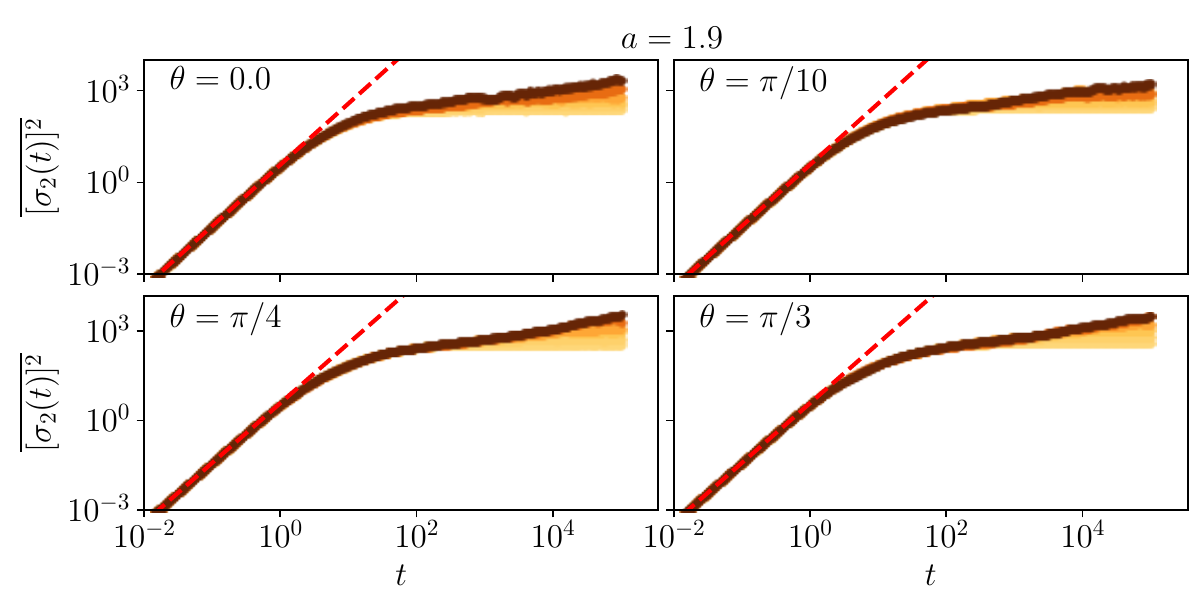}
\caption{\textit{Dynamics of the wave-packet width for $a=1.9$ and different values of $\theta$.} Compared to the earlier figures, the axes are not rescaled. The width becomes $N$-independent after the ballistic growth. Darkest color represents the largest system size, i.e., $N=8192$, and the data is averaged over 500 realizations for $N \leq 1024$ and 100 otherwise.}
    \label{fig:a-1.9}
\end{figure}

Finally, for $a=1.9$, shown in Fig.~\ref{fig:a-1.9}, we observe that after an initial perturbative ballistic growth, the wave-packet width becomes independent of the system size $N$. This saturation indicates the absence of long-time dynamics for $a>3/2$, consistent with the strongly localized nature of the eigenstates in this regime. It also agrees with the fac that $\gamma=0$ for $q=2,a>3/2$, as shown in sec \ref{app:sat-sigma-inf}.

\end{document}